\documentclass[12pt]{article}

\begin{document}

\title{Multidimensional quasi-exactly solvable potentials with two known eigenstates}
\author{V.M. Tkachuk, T.V. Fityo\\
  {\small Ivan Franko Lviv National University,
         Chair of Theoretical Physics }\\
        {\small 12 Drahomanov Str., Lviv UA--79005, Ukraine}\\
           {\small E-mail: tkachuk@ktf.franko.lviv.ua}}
\maketitle

\begin{abstract}
A general approach for constructing multidimensional quasi-exactly
solvable (QES) potentials with explicitly known eigenfunctions for
two energy levels is proposed. Examples of new QES potentials are
presented.


PACS number(s): 03.65.Ge
\end{abstract}
\maketitle

\section{Introduction}

The importance of exactly solvable potentials in quantum mechanics
is well known. But even in one-dimensional case the number of such
potentials is limited. Therefore, much attention has been given to
the quasi-exactly solvable (QES) pontentials for which  a finite
number of energy levels and corresponding wavefunctions are known
in explicit form. The first examples of such potentials were given
in \cite{aSiBiDu, aFl, aRa, aRa2, aKh} and subsequently many QES
potentials were established \cite{aTuUs, aTu, aTu2, aSh, aZaUlTs,
aZaUl, aUlZaVa, aGaKhSu} (for review see book by Ushveridze
\cite{aUs}).

In multidimensional case
many exactly solvable systems admit separation of variables and
thus they are reduced to one-dimensional problems \cite{Mi}. The
investigation of multidimensional and many-body systems which are
not amenable to separation of variables is important from
mathematical and physical points of view.

One of the most remarkable  exactly solvable models in $N$-body
quantum mechanics is the Calogero model \cite{aCa, aCa2}. This
model describes one-dimensional $N$-body problem with quadratic
and inverse square interacting potential. Soon after these papers
Sutherland extended Calogero model to a model where interaction
takes place on a circle \cite{aSu, aSu2, aSu3}. The further
progress was connected with Olshanetsky-Perelomov integrable
systems \cite{aOlPe} (see also review \cite{aOlPe2}). In
\cite{aMiRoTu} explicit examples of QES $N$-body problems on a
line were introduced for the first time. Later QES $N$-body
problems were studied in \cite{aHoSh, aBoLoTu}.

Recently, there has been achieved some progress in studying exact
solvability of such systems. In \cite{GhRa} a new class of QES
many-body Hamiltonians in arbitrary dimension was constructed. The
multidimensional Darboux transformation was proposed in
\cite{LoKa} and new examples of QES multidimensional matrix
Schr\"odinger operators were presented there. The authors of
\cite{UlLoRo} have developed a systematic procedure for
constructing exactly solvable and QES many-body potentials by
purely algebraic means. The method of multidimensional
supersymmetric (SUSY) quantum mechanics was applied to the
investigation of $N$-particle systems and an explicit construction
of exactly solvable 3-particle as well as QES $N$-particle
problems on a line  were presented \cite{IoNe, IoNe2, CaIo}. Very
recently, two new methods based on higher-order SUSY quantum
mechanics for the investigation of two-dimensional quantum
systems, whose Hamiltonians are not amenable to separation of
variables, were presented in \cite{CaIoNi}.

In this paper we develop a simple method for generation of QES
potentials with two known eigenstates in arbitrary dimension. For
one-dimensional case the problem of constructing QES potential with two
known levels was solved completely in the frame of SUSY quantum mechanics
\cite{Tk1, Tk2, Tk3} or using a simple method in which two wave functions
are chosen in such a way that they lead to the same potential
\cite{Ca, DoZa}. Just this method will be generalized for
multidimensional case.

Note that in fact the idea of the papers \cite{Tk1,Tk2,Tk3,Ca,
DoZa} is based on the inverse method. For the first time this
method was used by Ushveridze \cite{aUs} for construction QES
potentials.  It rests on the very simple idea: instead of looking
for solution of Schr\"odinger equation for a given potential one
should reconstruct the potential  starting with appropriately
chosen eigenfunctions. We would like to mention also paper
\cite{aBrDeNd} where detailed comparison of supersymmetric
approach proposed in \cite{Tk1} and Turbiner-Ushveridze approach
\cite{aTuUs} based on finite-dimensional representations of
$sl(2,R)$ was done.

\section{Construction of QES multidimensional potentials}

We consider the Schr\"odinger equation
\begin{equation}\label{e1}
H\psi=E\psi
\end{equation}
in $n$-dimensional case with the Hamiltonian
\begin{equation}\label{e2}
H=-\frac{1}{2}\Delta+V(x_1,\cdots x_n),
\end{equation}
where
$\Delta=(\mbox{\boldmath{$\nabla$}},\mbox{\boldmath{$\nabla$}})=\sum\frac{\partial^2}{\partial
x_i^2}$; $\mbox{\boldmath{$\nabla$}}$ is the nabla operator  in
$n$-dimensional space.

The wave function of the ground state with the energy $E_0$ has no
zeros and can be written in the  following form
\begin{equation}\label{e3}
\psi_0=e^{-F},
\end{equation}
where $F$ is a nonsingular function of $x_1,\dots x_n$.

Substituting  (\ref{e3}) into (\ref{e1}) we express the potential
in terms of the wave function $\psi_0$:
\begin{equation}\label{e4}
V=E_0+\frac{1}{2}\frac{\Delta\psi_0}{\psi_0}=E_0+\frac{1}{2}\left[
(\mbox{\boldmath{$\nabla$}} F)^2-\Delta F \right].
\end{equation}

Note that this transformation of Schr\"odinger equation from
linear form to non-linear one is well known in literature. See,
for instance, review by Turbiner, where this procedure was used
for construction of convergent perturbation theory in quantum
mechanics \cite{aTur}.

The wave  function of the excited state with the energy $E_1$ we write
as follows
\begin{equation}\label{e4s}
\psi_1=\phi e^{-F},
\end{equation}
where  $\phi=\psi_1/\psi_0$. This eigenfunction must lead to the
same potential $V$ and two eigenfunctions satisfy the following
equation
\begin{equation}\label{e5}
E_0+\frac{1}{2}\frac{\Delta\psi_0}{\psi_0}=
E_1+\frac{1}{2}\frac{\Delta\psi_1}{\psi_1}.
\end{equation}
In terms of $\phi$ and $F$ functions this equation reads
\begin{equation}\label{e5s}
2(\mbox{\boldmath{$\nabla$}}
F,\mbox{\boldmath{$\nabla$}}\phi)=\Delta\phi+2\varepsilon\phi,
\end{equation}
where
\[\varepsilon=E_1-E_0.\]

In one-dimensional case this equation can be easily solved with
respect to $F$ \cite{Ca, DoZa}. Then choosing different generating
functions $\phi$ we obtain different QES one-dimensional potentials
with two known eigenstates.

In multidimensional case the solution of equation (\ref{e5s}) is a
nontrivial task. In this paper we find the solution of (\ref{e5s})
with respect to $F$ for some special cases of function $\phi$.
Note that equation (\ref{e5s}) is a nonuniform linear equation
with respect to $F$. Any solution of it can be written in the
following form
\begin{equation}\label{e5d}
F=f+\tilde f,
\end{equation}
where $f$ is a particular solution of (\ref{e5s}) and $\tilde f$
is a general solution of the uniform equation
\begin{equation}\label{e5t}
(\mbox{\boldmath{$\nabla$}}\tilde
f,\mbox{\boldmath{$\nabla$}}\phi)=0.
\end{equation}
Direct substitution of function (\ref{e5d}) into equation
(\ref{e5s}) shows that $F$ is indeed a solution of the equation.
In order to satisfy square integrability of wavefunctions
(\ref{e3}) and (\ref{e4s}) we have to choose such $f$ and $\tilde
f$ that $F\to+\infty$ when $|\vec x|\to+\infty$.

Note that in one-dimensional case this equation gives only trivial
solution $\tilde f= {\rm const}$, which does not influence the
potential (\ref{e4}). New point of multidimensional case is that
uniform equation (\ref{e5t}) has many nontrivial solutions $\tilde
f$. This gives more possibilities for constructing QES potentials
in multidimensional case for given $\phi$ in comparison with the
one-dimensional case.

{\em Case 1.} This case corresponds to the following form of $\phi$
\begin{equation}\label{e6}
\phi=\sum_{i=1}^n\phi_i(x_i).
\end{equation}
Then equation (\ref{e5s}) can be rewritten as
\begin{equation}\label{e7}
2\sum_{i=1}^n{\phi_i}'\frac{\partial F}{\partial x_i}=
\sum_{i=1}^n\left({\phi_i}''+2\varepsilon\phi_i\right),
\end{equation}
where $\phi_i'=\frac{d\phi_i}{dx_i}$ and
$\phi_i''=\frac{d^2\phi_i}{d x_i^2}$.

The particular solution of equation (\ref{e7}) can be found in the
form of the function with separated variables
\begin{equation}\label{e8}
f=\sum_{i=1}^n f_i(x_i),
\end{equation}
where
\begin{equation}\label{e9}
\frac{df_i}{dx_i}=\frac{\phi_i''+2\varepsilon\phi_i+\lambda_i}{2\phi_i'},
\end{equation}
and the constants $\lambda_1,\dots\lambda_n$ have to satisfy the
following condition
\begin{equation}\label{e10}
\sum_{i=1}^n\lambda_i=0.
\end{equation}

The solution of the uniform equation
\begin{equation}\label{e11}
\sum_{i=1}^n\phi_i'\frac{\partial\tilde f}{\partial x_i}=0
\end{equation}
reads
\begin{equation}\label{e12}
\tilde f=\tilde f(\chi_1(x_1)-\chi_2(x_2),
\chi_1(x_1)-\chi_3(x_3),\dots,\chi_1(x_1)-\chi_n(x_n)),
\end{equation}
where $\tilde f$ is an arbitrary function of $n-1$ arguments and
\begin{equation}\label{e13}
\chi_i(x)=\int^x\frac{dy}{\phi_i'(y)}.
\end{equation}

Finally, the solution of (\ref{e7}) reads
\begin{equation}\label{e14}
F=\sum_{i=1}^n\int^{x_i}\frac {\phi_i''(x)+
2\varepsilon\phi_i(x)+\lambda_i} {2\phi_i'(x)}dx +\tilde
f(\chi_1(x_1)-\chi_2(x_2),\dots).
\end{equation}

Note that in the case of $\tilde f=0$ the variables of potential
(\ref{e4}) are separated. It is the function $\tilde f$ that is
responsible for nonseparability of variables in potential $V$.

{\em Case 2.} Let us choose the generating function $\phi$ in the
following form
\begin{equation}\label{e15}
\phi=\prod_{i=1}^n\phi_i(x_i).
\end{equation}

Now equation (\ref{e5s}) reads
\begin{equation}\label{e16}
2\sum_{i=1}^n\frac{\phi_i'}{\phi_i} \frac {\partial F} {\partial
x_i}=\sum_{i=1}^n\frac{\phi_i''}{\phi_i}+2\varepsilon.
\end{equation}
The solution is similar to the solution of the previous case. We
can write it as
\begin{equation}\label{e17}
F=\sum_{i=1}^n\int^{x_i}\frac {\phi''_i(x)+
(\frac{2}{n}\varepsilon+\lambda_i)\phi_i(x)}{2\phi_i'(x)}dx +\tilde
f(\chi_1(x_1)-\chi_2(x_2),\dots),
\end{equation}
where the constants $\lambda_1,\lambda_2,\dots$ also satisfy condition
(\ref{e10}), $\tilde f$ is an arbitrary function of $n-1$
arguments and
\begin{equation}\label{e18}
\chi_i(x)=\int^x\frac{\phi_i(y)}{\phi_i'(y)}dy.
\end{equation}

\section{Examples}

QES potential is given by equation (\ref{e4}), where the function $F$
is represented by expression (\ref{e14}) in the case 1 and by
expression (\ref{e17}) in the case 2. For this QES potential we know
two energy levels $E_0$ and $E_1$ as well as the corresponding
wave functions (\ref{e3}), (\ref{e4s}). Choosing various functions
$\phi_i$ for the case 1 or the case 2 we obtain different QES potentials.
Without loss of generality we choose in all expressions $E_0=0$
and then $E_1=\varepsilon$.

The functions $\phi_i$ and the parameters $\lambda_i$ must be
chosen in such a way that the function $f$ is a nonsingular one
and $\psi_0$, $\psi_1$ are square integrable functions. Note that
in multidimensional case we have also possibility to choose
different $\tilde f$ for this purpose.

{\em Example 1.} Let us choose
\[\phi=\frac{1}{2}\sum_{i=1}^na_ix_i^2
\]
which corresponds to the case 1. For this generating function we
obtain
\[\frac{df_i}{dx_i}=\frac{a_i+\lambda_i+a_i\varepsilon x_i^2}{2a_ix_i}.
\]
To satisfy the nonsingularity of the functions $f_i$ we have to
choose $\lambda_i=-a_i$. Note that $a_i$ satisfy the same
condition as $\lambda_i$, namely $\sum a_i=0$. Then
\[f_i(x_i)=\frac{\varepsilon}{4}x^2_i,\qquad\chi_i(x_i)=\frac{1}
{a_i}\ln x_i.\]

And finally, we obtain
\[F(x_1,x_2,\dots)=\frac{\varepsilon}{4}\sum_{i=1}^nx_i^2+\tilde
f\left(\frac{x_1^{1/a_1}}{x_2^{1/a_2}},\frac{x_1^{1/a_1}}{x_3^{1/a_3}}\dots\right)
,\]
here we rewrite
$\tilde f(\frac{1}{a_1}\ln x_1-\frac{1}{a_2}\ln x_2,
         \frac{1}{a_1}\ln x_1-\frac{1}{a_3}\ln x_3,\dots)$
as some new function
\[\tilde
f\left(\frac{x_1^{1/a_1}}{x_2^{1/a_2}},\frac{x_1^{1/a_1}}{x_3^{1/a_3}},\dots\right).\]

Let us apply these results to the two-dimensional case. Now,
$a_1+a_2=0$ and we may choose $a_1=-a_2=1$. Then
\[F(x,y)=\frac{\varepsilon}{4}(x^2+y^2)+\tilde f(xy).\]
QES potential and wave functions with zero and $\varepsilon$
energy levels read
\[V(x,y)=\frac{1}{2}\left[(\tilde f')^2-\tilde f''+
\frac{\varepsilon^2}{4}\right](x^2+y^2)+\varepsilon
xy\tilde f'-\frac{\varepsilon}{2},\]
\[\psi_0(x,y)=c_0e^{-\frac{\varepsilon}{4}(x^2+y^2)-\tilde f(xy)},\]
\[\psi_1(x,y)=c_1(x^2-y^2)e^{-\frac{\varepsilon}{4}(x^2+y^2)-\tilde
 f(xy)},\]
where $c_0$ and $c_1$ are normalization constants.

Choosing $\tilde f$ as some polynomial we reproduce the
two-dimensional potentials studied in \cite{Va} as interesting
examples of bottomless potentials with bound states.

For $\tilde f=0$ we obtain the isotropic harmonic oscillator.
Then, in Dirac notation, $\psi_0=|0,0\rangle$ is the ground state
eigenfunction, and
$\psi_1=\frac{1}{\sqrt{2}}(|2,0\rangle-|0,2\rangle)$ corresponds
to the second excited energy level.

Note that $\psi_0$ corresponds to the ground state if $\tilde f$
is a nonsingular function. For singular $\tilde f$ wave function
$\psi_0$ may
correspond to an excited state. For example, let us choose $\tilde
f=-\ln(xy)$. Then
\[\psi_0(x,y)=c_0xye^{-\frac{\varepsilon}{4}(x^2+y^2)}\]
corresponds to the second excited energy level.

{\em Example 2.} This example illustrates the case 2. We choose
\[\phi_i=x_i,\qquad \phi=\prod_{i=1}^n x_i.\]
Using (\ref{e17}) we obtain
\[F=\sum_{i=1}^n\left(\frac{2}{n}\varepsilon+\lambda_i\right) \frac{x_i^2}{4}+\tilde f(x_1^2-x_2^2,
x_1^2-x_3^2,\dots);\] remember that $\sum\lambda_i=0$.

Let us consider a particular three-dimensional case with
$\lambda_1=\lambda_2=\lambda_3=0$ and $\tilde
f=\alpha(2x^2-y^2-z^2)^2$. Then

\[V=\left(\frac{\varepsilon^2}{18}-16\alpha\right)x^2+\left
(\frac{\varepsilon^2}{18}-4\alpha\right)(y^2+z^2)+\]
\[+4\alpha(2x^2-y^2-z^2)^2\left(2\alpha(4x^2+y^2+z^2)+\frac{\varepsilon}{3}
\right)-\frac{1}{2}\varepsilon ,\]

\[\psi_0=c_0e^{-\varepsilon\frac{x^2+y^2+z^2}{6}-\alpha
(2x^2-y^2-z^2)^2},\]
\[\psi_1=c_1xyze^{-\varepsilon\frac{x^2+y^2+z^2}{6}-\alpha
(2x^2-y^2-z^2)^2},\] where $c_0$ and $c_1$ are normalization
constants.

For $\alpha=0$ we have harmonic oscillator. Then
$\psi_0=|0,0,0\rangle$ is the ground state function, and
$\psi_1=|1,1,1\rangle$ corresponds to the third excited energy
level. For $\alpha\ne 0$ we obtain nontrivial QES
three-dimensional potential, the variables of which cannot be
separated.

Here we would like to underline that question about separation of
variables in multidimensional Schr\"odinger equation is an
interesting and non-trivial task (see, for instance, \cite{aZh,
aZhZh}). We plan to discuss this problem in a part concerning our
approach in a separate paper.

\section{Conclusions}

We developed a simple approach for constructing QES
multidimensional potentials with two known energy levels and
corresponding wave functions. The proposed method is a direct
extension of the general approach proposed earlier for
one-dimensional case \cite{Ca,DoZa}. The central point of our
approach is equation (\ref{e5s}) and the main problem is to solve
this equation with respect to the function $F$. In contrast to
one-dimensional case, when the corresponding equation can be
easily solved, there is a nontrivial task to solve it in
multidimensional case. In this paper we find general solutions of
equation (\ref{e5s}) and construct new multidimensional QES
potentials for some special cases of function $\phi$. Finding of
other solutions of equation (\ref{e5s}) is an interesting problem
for further investigations.

Let us stress the following new point which appears in
multidimensional case in contrast to one-dimensional case. Namely,
in one-dimensional case $\tilde f={\rm const}$ and therefore the
ratio of the two eigenfunctions $\psi_1/\psi_0=\phi$ and distance
between corresponding energy levels $\varepsilon$ entirely
determine the potential energy $V$ \cite{Tk2, Ca, DoZa}. In
multidimensional case the potential energy $V$ is determined by
$\phi$, $\tilde f$ and $\varepsilon$. Equation (\ref{e5t}) allows
many nontrivial solutions for $\tilde f$. As result we obtain a
family of potentials $V$ for fixed $\phi$ and fixed energy gap
$\varepsilon$ choosing different solution for $\tilde f$. This
feature of multidimensional case is explicitly shown in the
example 1, where potential $V(x,y)$ for fixed $\phi$ and
$\varepsilon$ depends on arbitrary function $\tilde f(xy)$. Thus,
in multidimensional case there are more possibilities for
constructing QES potentials for a given $\phi$ in comparison with
the one-dimensional case.

In one-dimensional case all eigenfunctions can be easily ordered
using oscillation theorem which states that the $n$-th
eigenfunction has $n$ zeros. For multidimensional case it is known
that ground state eigenfunction also has no zero and the
corresponding energy level is non-degenerated \cite{ReSi}, but
there is no theorem which connects zeros and order of excited
eigenfunctions. Therefore, one cannot give an exact answer between
which states the energy gap is calculated. We can only state that
if an eigenfunction has no zeros then this function corresponds to
the ground state.

\end{document}